\documentclass[sigconf,authorversion]{acmart}

\usepackage{mathtools}
\usepackage{amsmath}

\usepackage{pgfplots}
\usepackage{graphicx}
\usepackage{wrapfig}
\usepackage{amsthm}
\usepackage{tikz}
\usepackage{tikz-cd}
\usepackage{enumerate}
\usepackage{booktabs}

\usepackage{float}
\usepackage{natbib}

\pgfplotsset{width=10cm,compat=1.9}
\usepackage{natbib}
\usepackage{hyperref}
\hypersetup{
    colorlinks=true,
    linkcolor=blue,
    citecolor=blue
}
\usepackage[capitalize]{cleveref}
\usepackage{caption}
\usepackage{subcaption}

\usetikzlibrary{positioning,chains,fit,shapes,calc}

\newtheorem{theorem}{Theorem}
\newtheorem{lemma}[theorem]{Lemma}
\newtheorem{proposition}[theorem]{Proposition}

\newtheorem{definition}[theorem]{Definition}
\newtheorem{assumption}[theorem]{Assumption}

\newcommand{\pp}{\mathbb{P}}
\newcommand{\F}{\mathcal{F}}

\newcommand{\abs}[1]{\left| #1 \right|}

\newcommand{\ee}{\mathbb{E}}

\newcommand{\R}{\mathfrak{R}}

\newcommand{\qbar}{\overline{q}}
\newcommand{\pqa}{p_{\rank_q(a)}}
\newcommand{\pqba}{p_{\rank_{\qbar}(a)}}

\DeclareMathOperator*{\argmin}{argmin}
\DeclareMathOperator{\variance}{Var}
\renewcommand{\epsilon}{\varepsilon}
\renewcommand{\phi}{\varphi}
\DeclareMathOperator{\rel}{rel}
\DeclareMathOperator{\rank}{rank}
\DeclareMathOperator*{\argmax}{argmax}
\DeclareMathOperator{\sgn}{sign}

\copyrightyear{2022} 
\acmYear{2022} 
\setcopyright{rightsretained} 
\acmConference[SIGIR '22]{Proceedings of the 45th International ACM SIGIR Conference on Research and Development in Information Retrieval}{July 11--15, 2022}{Madrid, Spain}
\acmBooktitle{Proceedings of the 45th International ACM SIGIR Conference on Research and Development in Information Retrieval (SIGIR '22), July 11--15, 2022, Madrid, Spain}\acmDOI{10.1145/3477495.3531958}
\acmISBN{978-1-4503-8732-3/22/07}
\begin{document}

\title[Utility-Aware Ranking for Query Autocomplete]{Counterfactual Learning To Rank for Utility-Maximizing Query Autocompletion}

\author{Adam Block}
\authornote{Work done while at Amazon}
\email{ablock@mit.edu}
\affiliation{%
  \institution{Massachusetts Institute of Technology}
  \city{Cambridge}
  \state{Massachusetts}
  \country{USA}
}
\author{Rahul Kidambi}
\email{rahul.g.kidambi@gmail.com}
\author{Daniel N. Hill}
\email{daniehil@amazon.com}
\affiliation{%
    \institution{Amazon Search}
    \city{Berkeley}
    \state{California}
    \country{USA}
}
\author{Thorsten Joachims}
\email{thorstj@amazon.com}
\affiliation{%
    \institution{Amazon Music} 
    \city{San Francisco}
    \state{California}
    \country{USA}
}
\author{Inderjit S. Dhillon}
\email{inderjit@cs.utexas.edu}
\authornotemark[1]
\affiliation{%
    \institution{UT Austin}
    \city{Austin}
    \state{Texas}
    \country{USA}
}

\begin{abstract}
  Conventional methods for query autocompletion aim to predict which completed query a user will select from a list. A shortcoming of this approach is that users often do not know which query will provide the best retrieval performance on the current information retrieval system, meaning that any query autocompletion methods trained to mimic user behavior can lead to suboptimal query suggestions. To overcome this limitation, we propose a new approach that explicitly optimizes the query suggestions for downstream retrieval performance. We formulate this as a problem of ranking a set of rankings, where each query suggestion is represented by the downstream item ranking it produces. We then present a learning method that ranks query suggestions by the quality of their item rankings. The algorithm is based on a counterfactual learning approach that is able to leverage feedback on the items (e.g., clicks, purchases) to evaluate query suggestions through an unbiased estimator, thus avoiding the assumption that users write or select optimal queries. We establish theoretical support for the proposed approach and provide learning-theoretic guarantees. We also present empirical results on publicly available datasets, and demonstrate real-world applicability using data from an online shopping store.
\end{abstract}

\maketitle
\section{Introduction}\label{sec:intro}

Query autocompletion (QAC) systems~\citep{Bar-YossefK11,Shokouhi13,CaiR16c,SordoniBVLSN15,DBLP:journals/corr/abs-2012-07654} recommend candidate query completions given a partially typed query, and QAC has become a standard feature of many retrieval systems that are currently in practical use. We argue that the goal of QAC is not only to reduce the user's typing effort, but also to help users discover the best queries for their information needs. Unfortunately, there is a disconnect between how most current QAC systems are trained and the ultimate goal of finding improved queries. In particular, most QAC systems are trained to mimic user behavior, either predicting how the user will complete the query or which query suggestion the user will select. This means that a conventional QAC system can only become as good at suggesting queries as the users it was trained on, which may lead to substantial suboptimality.

To overcome this limitation, we present a new framework for training QAC systems, which we call the utility-aware QAC approach. The name reflects that the QAC system is aware of the utility (i.e. ranking performance) that each query suggestion achieves given the current production ranker, and that it directly optimizes the retrieval performance of the queries it suggests. The key insight is to make use of downstream feedback that users eventually provide on the items (e.g. products purchased and content streamed) to evaluate each suggested query instead of considering previous users' queries as the gold standard to emulate.  This new focus allows our approach to circumvent the issue that users may not know which queries provide good rankings given the current system.

From a technical perspective, the utility-aware QAC approach formulates the task of learning a QAC system as that of learning a ranking of rankings. In particular, each query suggestion is evaluated by the quality of the item ranking it produces. The goal of the utilitiy-aware QAC is to rank the query suggestions by the quality of their item rankings given the partial query that the user already typed. A key machine learning challenge lies in how to estimate the quality of the item rankings of each of the query suggestions in the training set, given that we only have access to interaction feedback (e.g. purchases) for a small subset of query suggestions and items. We overcome this problem by taking a counterfactual  learning approach, where we develop an estimator of ranking performance that is unbiased in expectation under standard position-bias models \cite{joachims2017unbiased}. This results in a new training objective for QAC models that directly optimizes the efficacy of the queries suggested by the QAC system.  Note that at test time, the system will not have access to any utility estimate as the user's desired document is obviously not known; the goal is for the ranker to use the interaction of features relevant to the user (such as a prefix or contextual data) and the query to predict the downstream utility of different queries, and then to surface high quality suggestions to reduce downstream user effort.  Thus, it is critical for the utility-awareness of the proposed framework to incorporate the downstream effect somewhere in the objective, as we do, and thus not rely on access to a utility estimate at test time.

We now list our primary contributions.  
\begin{itemize}
    \item We introduce a novel framework for training utility-aware QAC systems given biased, item-level feedback.  In particular, we propose a realistic theoretical model and a learning method that can train a utility-aware QAC system given an arbitrary class of potential ranking functions.
    \item We provide a theoretical analysis and show that under mild conditions on the function class used for QAC ranking and for a known position-bias model with full support, our approach to training QAC rankers is consistent in the sense that it will identify the best-in-class QAC ranker given sufficient training data.  We also state and prove a non-asymptotic anologue of this result.
    \item Finally, we investigate empirically how well the utilitiy-aware QAC approach performs by instantiating our method, both on public benchmarks and on a real-world dataset from an online shopping store. The latter demonstrates real-world practicality, while the former gives insight into how various features of the utilitiy-aware QAC approach contribute to improved efficacy.
\end{itemize}
We emphasize that our proposed framework is invoked only at training time and thus does not affect latency at inference time.  In particular, our framework naturally adapts to essentially any QAC approach and can scale to extremely large (tens of millions) query and document universes and thus can realistically be deployed in many practical settings.  The structure of the paper is as follows.  We first provide a brief survey of related work.  We then formally propose the utility-aware QAC framework as a ``ranking of rankings'' and continue by proposing an unbiased estimator of utility, given possibly biased data.  We proceed by stating a nonasymptotic generalization bound for a ranker trained in our framework, which in turn implies consistency.  We then move on to describe the practical instantiation of our framework along with a description of the experimental setup.  We conclude by presenting the results of our experiments.  All proofs are deferred to \cref{app:proofs}.

\subsection{Related Work}
{\bf Query auto-complete system}: QAC has been studied extensively in the literature - particular efforts include suggesting top-k queries given a prefix~\citep{WangGGL20} and contextual, personalized, time-sensitive and diversified recommendation of query completions for real time applications~\citep{Bar-YossefK11,ShokouhiR12,Shokouhi13,CaiLR14,CaiLR16,DBLP:journals/corr/abs-2012-07654,CaiRR16b}. See~\citep{CaiR16c} for a survey of these methods. Common to the above approaches is the fact that they work in a two-stage retrieve and rank framework, where, given a prefix, a candidate set of query completions is retrieved and then re-ranked using context, popularity and other metrics, and the resulting top-$k$ queries are shown to the user. Techniques from eXtreme multi-label learning~\citep{DBLP:journals/corr/abs-2012-07654} have also been applied to retrieval for QAC systems \citep{YenHRZD16,MineiroK14a,PrabhuKHAV18,yu2020pecos}. These approaches optimize for user engagement measured in the form of clicks on the presented queries. This line of work is different from the goals of this paper in that we minimize downstream user effort as opposed to maximizing user engagement.  Another class of QAC approaches include ones based on neural language models, performing generation/re-ranking~\citep{DehghaniRAF17,ParkC17,JaechO18,SordoniBVLSN15}. However, these approaches may not be suitable to real-time deployment owing to latency issues and their propensity to offer non-sensical suggestions in the generative setting.

{\bf Ranking in full and partial information settings}: Ranking in the full information setting has been studied extensively, including extensions to partial relevance judgements \cite{joachims02}. For a survey of detailed developments in the pairwise learning to rank framework, see \cite{Burges10}. This line of work assumes the relevance judgements can be used to optimize metrics of interest. Recognizing the biased nature of feedback received by data collected through deploying a ranking algorithm, \cite{joachims2017unbiased} developed a de-biased learning to rank approach with provable guarantees; this method then inspired a plethora of extensions, generalizations, and applications to other domains such as \cite{Chen:2016:XST:2939672.2939785,wang2018position,yang2018unbiased}. We employ a similar debiasing strategy for purposes of estimating utilities in order to develop an unbiased utilitiy-aware learning to rank approach. 

{\bf Counterfactual estimation/reasoning}: At a high level, this work is motivated by principles of counterfactual estimation/learning with logged bandit feedback developed by \cite{bottou2012}. The utilitiy-aware ranking problem is related to the general estimation/learning from logged contextual bandit feedback framework that has been studied in a series of papers \cite{Horvitz52,Strehl10,Dudik11,SwaminathanJ15,WangAD16,SwaminathanKADL16,Joachims18,Farajtabar18,Su20}.

\section{Problem Setup}\label{sec:problem_setup}
Before providing the formal setup, we first need to review common notions from the literature (see \cite{joachims2017unbiased} for more details).  In standard ``full information'' Learning to Rank (LTR), the learner desires a policy that maps from queries $\mathcal{Q}$ to a ranked list of documents $\mathcal{A}$, attempting to place documents relevant to the query near the top of the list.  Relevance is measured by a score $\rel(q, a)$ that is assumed to be known during training but not at inference time.  In the sequel, for the sake of simplicity, we restrict our focus to relevances in $\{0,1\}$ but we note that our techniques are applicable to real valued relevances.  In the context of QAC systems, the `documents' are query completions which are ranked according to their relevance to a context, such as a prefix.  The quality of a given document ranker, denoted by $\rank$, evaluated at a query $q$ is measured by an additive utility function $\Delta(q, \rank)$; examples include Mean Reciprocal Rank (MRR) and Discounted Cumulative Gain (DCG) \cite{jarvelin2002cumulated}.  To evaluate the ranker, it is common to consider the value of $\Delta(q, \rank)$ averaged over some distribution of queries.

In contradistinction to the traditional LTR setting, our goal is to produce a \emph{ranking of rankings} given a context, where the learner associates an element of a set of pre-determined rankings to each context.  Thus, we consider a space of contexts $\mathcal{X}$ that represent partial queries, a universe of query completions $\mathcal{Q}$, and set of documents $\mathcal{A}$.  Crucially, in our setting, there is a given document ranker that maps a query to a ranked list of documents:
\begin{definition}\label{as1}
    There exists a fixed function $\rank : \mathcal Q \to (\mathbb N \cup \{\infty\})^{\mathcal A}$ that acts as a ranker of documents given a query, i.e., $\rank_q$ is a function mapping articles to ranks.  We denote by $\rank_q(a)$ the rank of document $a$ given query $q$ and consider $\rank_q(a) = \infty$ to suggest that the query $q$ does not return the document $a$.  By abuse of notation, we also denote the set of documents returned by a query, $\{a | \rank_q(a) < \infty \}$ by $q$ when there is no risk of confusion.
\end{definition}

Our goal is to produce a query ranking policy, $S: \mathcal{X} \to (\mathbb{N} \cup \{\infty\})^{\mathcal{Q}}$ that highly ranks queries leading to contextually relevant documents with minimal effort.  Thus, we are concerned with relevances  $\rel(x,a)$ between contexts and documents and, given a context $x$, are attempting to find queries $q$ such that $\rank_q(a)$ is small for documents $a$ when $\rel(x, a)$ is large.  With full information on all relevance labels, where the learner has offline access to  $\Delta(q, \rank)$, a natural approach to the ranking-of-rankings problem would be to define relevance between context and query as $\rel(x, q) = \Delta(q, \rank)$.  Then, given a (possibly different) additive utility function $\widetilde{\Delta}$, we can quantify the quality of a query ranker $S$ evaluated on a context $x$ using these scores.  Given a distribution of contexts, we evaluate the query-ranking policy as follows:
\begin{equation}\label{eq:reward}
    L(S) = \ee\left[\widetilde{\Delta}(x, S)\right]
\end{equation}
Thus, we see that the ranking of rankings problem with full information can be reduced to LTR.  In the QAC setting, the context $x$ is data associated to a user, such as a typed prefix while relevances are often measured by clicks.  We also suppose the learner has offline access to the fixed document retrieval system from \cref{as1}.  Given that the QAC and the document retrieval system are trained separately, it is reasonable to take the latter as a black box that can be queried offline.
\begin{definition}\label{as1b}
    An offline ranking system returns $\rank_q(a)$ given a $(q, a)$ pair.
\end{definition}
While \cref{as1b} allows the learner to query the document ranker, one couldn't hope to do this during inference time (owing to latency constraints), thus requiring a reliance on a pre-trained ranker, $S$.  Much as in traditional LTR, we aim to choose an $S$ maximizing $L(S)$, for which we use an empirical proxy:
\begin{equation}
    \widehat{L}(S) = \frac 1n \sum_{i = 1}^n \widetilde{\Delta}(x_i, S)
\end{equation}
The law of large numbers tells us that $\widehat{L}$ converges to $L$ as $n \to\infty$ under the following generative process:
\begin{definition}\label{as2}
    Data are generated such that we receive $n$ contexts $x_i$ sampled independently from a fixed population distribution.  
 \end{definition}
Given a function class $\F$ of candidate rankers, we can optimize $\widehat{L}$ instead of $L$, leading to the classically studied Empirical Risk Minimizer (ERM).  Convergence rates of $\widehat{S}$ to the optimal ranker $S^\ast$ (that which maximizes $L$) can then be established depending on the complexity of the function class $\F$ on the basis of empirical process theory \cite{wainwright2019high}.  This analysis, however, is predicated on the assumption that $\widehat{L}(S)$ can be evaluated, which in turn requires known relevances $\rel(x, a)$ between documents and contexts.  As noted in \cite{joachims2017unbiased}, even in the LTR setting, collecting such relevance scores can be challenging, commonly generated by human judges making potentially inconsistent determinations.  To escape these difficulties, we use logged data, as elaborated below.

\section{Ranking of Rankings with Partial Information}\label{sec:partial_info}
In the previous section, we saw that the ranking of rankings task in a full-information setting can be reduced to the standard LTR task, but noted that acquiring reliable data in this regime presents its own challenges.  In standard LTR, \cite{joachims2017unbiased} proposed using logged data to estimate relevance between queries and documents.  As many services keep data on user searches and clicks, these logs provide a rich source of information on what documents the customers themselves consider to be relevant to their context.  Despite the quality and quantity of the log data, their use to measure relevance still requires care as relevance observed by the learner is a function both of relevance between context and document and, critically, whether or not the customer associated to the context actually observed the document.  Following \cite{Schnabel2016UnbiasedCE,joachims2017unbiased}, we assume the probability of observation is dependent on the rank and we apply inverse propensity scoring to produce an unbiased estimate of $L$.

More formally, we consider the following model.  Given query $q$ and document $a$, we denote the event that document $a$ was observed by the customer given query $q$ by $o(q, a)$.  Given a context $x$, we denote by $c(x,q, a)$ the event that a document is \emph{clicked} (and its relevance logged) and suppose this occurs if and only if the document is observed and the document is relevant, i.e., $c(x, q, a) = o(q, a) \rel(x, a)$.  We suppose the following distribution of $o(q,a)$:
\begin{definition}\label{as3}
    The data generation process from log data has the events $\{o(q, a)| q \in \mathcal Q, \,\, a \in \mathcal A\}$ live on a probability space such that the events are independent across queries.  We further assume that there exists a known probability distribution $\{p_k\}$ on $\mathbb{N} \cup \{\infty\}$ such that $\pp(o(q, a) = 1) = p_{\rank_q(a)}$ for all $q, a$, where $p_\infty = 0$.
\end{definition}
The most restrictive part of \cref{as3} is the assumption of the known click propensity model.  Such a model can be estimated in various ways \cite{10.1145/2911451.2911537,joachims2017unbiased}, and we do not concern ourselves with specifying a particular model.  In our experiments below, we consider a simple model where $p_k \propto \frac 1k$; furthermore, we demonstrate that the proposed framework can offer gains even in situations when the propensities are mis-specified. 

We focus on the most natural reward for a query given a context: the probability of a click.  Thus we define the utility of a query as
\begin{align}\label{eq:utility}
    u(x, q) &= \pp(c(x, q, a) = 1 \text{ for some } a \in q) 
\end{align}
The learner aims to produce a query ranker, which takes in a context and returns a ranking of queries.  Thus if 
\begin{equation}
    q^* (x) \in \argmax_{q \in Q(x)} u(x, q)
\end{equation}
the goal is for the query ranker to rank near the top queries $q \in Q(x)$ such that $u(x, q)$ is as close as possible to $u(x, q^\ast(x))$.

\section{Counterfactual Utility Estimation}\label{sec:utilityEstimation}
If we knew the relevances of all context-document pairs, we would have access to $u(x, q)$ and could proceed as described in Section \ref{sec:problem_setup}.  Because we only have access to log data, we need to form an estimate of the utility of different queries given the data at hand.  The following describes what we need from the logs:
\begin{definition}\label{as4}
    Let $Q$ denote a pre-trained function mapping a context $x$ to the set of proposed queries.  A data point in the logs $\left(x_i, \qbar_i, a_i, \rank_{\qbar_i}(a_i)\right)$ consists of context $x$
    , query $\qbar_i \in Q(x_i)$ chosen arbitrarily, document $a_i \in \qbar_i$ such that $c(x, \qbar, a) = 1$, and $\rank_{\qbar}(a_i)$. 
\end{definition}
Given a data point $(x, \qbar, a, \rank_{\qbar}(a))$, we consider the following estimator of utility for all $q \in Q(x)$:
\begin{equation}
    \widehat{u}(x, q| \qbar, a) = \sum_{\substack{a \in q \\ c(x, \qbar, a) = 1}} \frac{p_{\rank_q(a)}}{p_{\rank_{\qbar}(a)}}
\end{equation}
We first note that our estimator, motivated by the Inverse Propensity Scoring (IPS) of \cite{joachims2017unbiased}, is unbiased.
\begin{proposition}\label{prop:unbiased}
    Suppose we are in the setting of Definitions \ref{as1}, \ref{as1b}, \ref{as2}, \ref{as3}, and \ref{as4}.  Then $\widehat{u}$ is an unbiased estimator of the utility, i..e, $\ee\left[\widehat{u}(x, q| \qbar, a) | x, q\right] = u(x, q)$.
\end{proposition}

Given that our estimator is unbiased, we might now hope to control its variance.  Unfortunately, as the next result shows, this is not possible without further assumptions:
\begin{proposition}\label{prop:novariancecontrol}
    For any constant $C$, there exist queries $q$ and $\overline{q}$, a context $x$, and document $a$ such that Assumptions \ref{as1}, \ref{as1b}, \ref{as2}, \ref{as3}, and \ref{as4} hold and $\variance(\widehat{u}(x, q| \qbar, a)) > C$.
\end{proposition}
This makes intuitive sense: there is no reason that it should be easy to estimate the utility of a query $q$ if we only have data from a query $\qbar$ that is very different from $q$. Thus, in order to estimate utilities well, we need to control this difference.  The following assumption controls this difference quantitatively:
\begin{assumption}\label{as5}
    We assume that there is a positive constant $B < \infty$ such that
    \begin{equation}
        \sup_x \max_{q, \qbar \in Q(x)} \max_{r(x, a) = 1} \frac{\pqa}{\pqba} \leq B
    \end{equation}
\end{assumption}
It is important to note that \cref{as5} makes no reference to the probability ratio with respect to irrelevant documents. Thus, $q$ and $\qbar$ can be arbitrarily different in ranking irrelevant documents.  This is similar to the full coverage assumptions used in the off-policy evaluation and learning literature~\citep{Strehl10,Dudik11,WangAD16}. Note that such a bound on the ratio of probabilities can be enforced through clipping weights~\citep{Strehl10} which reveals a bias-variance tradeoff when running the resulting estimation procedure. Owing to this, \cref{as5} is not very restrictive, as practitioners often treat documents as irrelevant if they appear in the logs only with very large ranks.  Under \cref{as5}, and using Equation \eqref{eq:var2} we are able to control our estimator's variance:
\begin{proposition}\label{prop:variance}
    Suppose we are in the setting of Definitions \ref{as1}, \ref{as1b}, \ref{as1b}, \ref{as2}, \ref{as3}, and \ref{as4}, as well as Assumption \ref{as5}.  Then
    \begin{align}
        \widehat{u}(x, q| \qbar, a) &\leq B \label{eq:varbound1} \\
        \variance(\widehat{u}(x, q| \qbar, a)) &\leq B u(x, q) - u(x, q)^2 \label{eq:varbound2}
    \end{align}
\end{proposition}
Our utility estimator gets better the closer that $q$ is to $\qbar$ (in terms of the rankings they produce on relevant documents).  
We now proceed to prove a generalization bound.

\section{Generalization Under Partial Information}\label{sec:Generalization}
In this section, we state a generalization bound for our ranking of rankings in the above model.  We restrict our focus to a variant of the Pairwise Learning to Rank model \cite{herbrich,joachims02}, where we are given a dataset containing groups of features and targets.  Within each group, we subtract the features of all pairs with different target values, and label this difference $1$ if the first target is larger than the second and $-1$ otherwise.  We then train a classifier, such as a logistic regression model, on these data to predict the assigned labels.  To produce a ranking given features, we call the classifier, which returns a real number between 0 and 1, interpreted as a probability that one is better than the other and then rank the candidates by their predicted probabilities, within group.

For the sake of simplicity, we focus our theoretical analysis on the problem of minimizing average loss in utility obtained by transposing any two elements in the ranking.  This metric is particularly well suited to theoretical analysis with respect to PLTR and is a natural generalization to the problem of ranking more than two items while keeping target score values relevant to the loss considered in such works as \cite{clemenccon2008ranking,JMLR:v6:agarwal05a}, among others.  

Given query ranker $S: \mathcal{X} \to (\mathbb{N} \cup \{\infty\})^{\mathcal Q}$, for a context $x$, queries $q, q' \in Q(x)$, let $S(x, q, q') = -\sgn\left(\rank_{S(x)}(q) - \rank_{S(x)}(q') \right)$ with $\sgn(0)$ chosen arbitrarily, where we denote by $\rank_S(q)$ the rank of query $q$ according to the ranker $S$; because there is no overlap between contexts, queries, and documents, there is no risk of confusion with the document ranker defined earlier.  In other words, $S(x, q, q')$ is $1$ if $S(x)$ ranks $q$ ahead of $q'$ and $-1$ otherwise.  We then formally define the loss of a query ranker to be:
\begin{equation}\label{eq:classifier}
    \widetilde{\Delta}(x, S) = \frac 1{\binom{\abs{Q(x)}}{2}} \sum_{q, q' \in Q(x)} (u(x, q) - u(x, q'))S(x, q, q')
\end{equation}
Taking expectations yields $L(S)$.
As noted in Section \ref{sec:partial_info}, we don't have access to utilities since we work in the partial information setting; instead we have access to (relative) utility estimates $\widehat{u}(x, q| \qbar, a)$.  Thus, we consider:
\begin{equation}
   \widetilde{L}(S) = \ee\left[\frac 1{\binom{\abs{Q(x)}}{2}} \sum_{q, q' \in Q(x)} (\widehat{u}(x, q| \qbar, a) - \widehat{u}(x, q'| \qbar, a))S(x, q, q')\right] 
\end{equation}
The tower property of expectations, linearity, and \cref{prop:unbiased} show that the difference between these losses is merely notational:
\begin{lemma}\label{lem:losses}
    Under the setting of Definitions \ref{as1}, \ref{as1b} \ref{as2}, \ref{as3}, and \ref{as4}, for any query ranker $S: \mathcal{X} \to (\mathbb{N} \cup \{\infty\})^{\mathcal Q}$, we have $\widetilde{L}(S) = L(S)$.
\end{lemma}

Thus, minimizing $L$ is the same as minimizing $\widetilde{L}$; unfortunately, we don't have access to $\widetilde{L}$ either as we don't know the population distribution.  We consider the empirical version, $\widehat{L_n}(S)$, defined as:
\begin{align}
    \frac 1n \sum_{i = 1}^n \frac 1{\binom{\abs{Q(x_i)}}{2}} \sum_{q, q' \in Q(x_i)} (\widehat{u}(x_i, q| \qbar_i, a_i) - \widehat{u}(x_i, q'| \qbar_i, a_i))S(x_i, q, q')
\end{align}
The learner has access to $\widehat{L}_n$ and can thus optimize over some class of rankers $\F$.  We consider $S_n \in \argmin_{S \in \F} \widehat{L}_n(S)$, the ERM.  We wish to analyze the difference in performance between $S_n$ and the best ranker in the class $\F$, i.e., we hope that $L(S_n) - L(S^*)$ is small, where $S^* \in \argmin_{S \in \F} L(S)$.   Classical theory of empirical processess \cite{devroye2013probabilistic,wainwright2019high} suggests that generalization error depends on the complexity of the function class, through the Rademacher complexity.  Letting $\F' = \{S(x, q, q')| S \in \F\}$ where $S(x, q, q')$ is as in \cref{eq:classifier}, we let the Rademacher complexity be defined as
\begin{equation}
    \R_n(\F) = \ee\left[\sup_{S \in \F'} \frac 1n \sum_{i = 1}^n \epsilon_i S(x_i, q_i, q_i') \right]
\end{equation}
where the $\epsilon_i$ are independent Rademacher random variables, the $x_i$ are chosen independently according to the distribution on $\mathcal X$, and the $q_i,q_i'$ are a fixed set of queries.  As a concrete example, it is easy to show that if 
$\F$ is linear, then $\R_n(\F) = O\left(d n^{- 1/2}\right)$ with a constant depending on the norm of the parameter vector.  There is a great wealth of classical theory on controlling $\R_n(\F)$ in many interesting regimes and we refer the reader to \cite{wainwright2019high,devroye2013probabilistic,van2014probability} for more details.  By adding queries with zero utility, we may suppose that $\abs{Q(x)} = K$ for all $x$. 
We now state the generalization bound:
\begin{theorem}\label{thm:generalization}
    Suppose definitions \ref{as1}, \ref{as1b} \ref{as2}, \ref{as3}, \ref{as4}, and \cref{as5} holds.  Let $\F$ be any class of query rankers $S: \mathcal X \to (\mathbb{N} \cup \{\infty\})^{\mathcal{Q}}$.  Then we can control the generalization error as follows:
    \begin{equation}
        \ee\left[L(S_n) - L(S^\ast)\right] \leq 4 B \left(\frac{\rho_n}{K}\right)^2 \R_n(\F) = O\left(\frac{1}{\sqrt n}\right)
    \end{equation}
    where the expectation is taken with respect to the data used to construct $S_n$, $\rho_n$ is the expected number of queries relevant to at least one $x_i$ for $1 \leq i \leq n$, and the equality holds for parametric function classes $\F$.
\end{theorem}
As discussed in the previous section, a linear dependence on $B$ is unavoidable in the absence of further assumptions.  The quantity $\rho$ is obviously bounded by $\abs{\mathcal Q}$ but can in general be much smaller if, for most contexts, only a small number of queries are relevant.  Without further structural assumptions on the query universe, it is impossible to escape such a dependence as there is no ability to transfer knowledge about one set of queries to another.  
From the bound, it seems like increasing $K$ can only be advantageous, but this is not quite true because increasing $K$ weakens the loss function's ability to distinguish queries ranked near the top; for practical applications, users rarely look sufficiently far down a ranked list for large $K$ to be relevant.  
What is clear from this bound, however, is that our approach is consistent and utility-aware ranking with partial information is well-grounded in theory.

\section{Practical Instantiation}\label{sec:pracInstantiation}
We discuss the practical instantiation of the principles developed in Sections~\ref{sec:partial_info}-\ref{sec:Generalization}.  We first present the system architecture and then describe the setup used to evaluate the efficacy of our approach.

\subsection{System Architecture}\label{ssec:sysArch}
For computational reasons, the QAC pipeline involves two steps \cite{DBLP:journals/corr/abs-2012-07654}: we first retrieve a set of possible query completions and second re-rank this set.  Both steps are utility-aware.

\subsubsection{Query Retrieval}\label{sssec:qRetrieval}
Selecting a small subset of possible query completions is beneficial during both training and inference.  For training, this is related to the PLTR reduction, described above, where the number of samples for the classification task grows quadratically with the number of queries per context. Moreover, owing to latency requirements (typically around 100 milliseconds) of practical QAC systems with query universes ranging from 100K to 100M, evaluating all possible queries for a given prefix is impractical.  Thus, logarithmic time query retrieval strategies can be achieved in practice by relying on techniques from eXtreme multi-label learning~\citep{YenHRZD16,MineiroK14a,PrabhuKHAV18,yu2020pecos} and maximum inner product search~\citep{CremonesiKT10,DeanRSSVY13,WangSSJ14,Shrivastava014b,GuoSLGSCK20}, among others.

In this work, we apply Prediction for Enormous and Correlated Output Spaces (PECOS) \citep{yu2020pecos}, proposed in the context of eXtreme multi-label learning, to perform candidate query retrieval. Building on~\citep{DBLP:journals/corr/abs-2012-07654}, we use hierarchical clustering to index queries and train linear classifiers for each node in the tree; PECOS then uses beam search to retrieve relevant query completions given the context.  To make the retriever utility-aware, we include as relevant to a context only sufficiently high-utility queries as training data.

\subsubsection{Query Re-Ranking}\label{sssec:qReranking} 
While the query retrieval system optimizes recall so as to not exclude optimal queries, re-ranking is required to improve precision at the top ranks.  There are many ranking metrics \citep{Joachims05a,KarN014} and algorithms based on gradient boosting and deep neural networks with the PLTR objective have been widely adopted in practice \citep{Burges05,Burges10,ChenG16,YZTPWBN21}.  In this work, we use the former, implemented in the library \textsf{xgboost} \citep{ChenG16}.

\subsection{Experimental Setup}\label{ssec:exptDesign}
Given the log data, we first partition the training set in three parts, with one set used to train the query retriever, the second used to train the query ranker and the third used for evaluating the entire pipeline.  In order to train the retriever model, we must feed it proposed alternative queries.  To generate these candidates, we use the log data itself, along with the document ranker.

\definecolor{myblue}{RGB}{80,80,160}
\definecolor{myred}{RGB}{160,80,80}
\definecolor{mygreen}{RGB}{80,160,80}
\begin{figure}[ht]
\centering
\scalebox{.7}{
\begin{tikzpicture}[thick,
  every node/.style={draw,circle},
  fsnode/.style={fill=myblue},
  ssnode/.style={fill=myred},
  every fit/.style={ellipse,draw,inner sep=-2pt,text width=2cm},
  -,shorten >= 3pt,shorten <= 3pt
]

\begin{scope}[start chain=going below,node distance=7mm]
\foreach \i in {1,2,...,4}
  \node[fsnode,on chain] (f\i) [label=left: $\mathbf{q_\i}$] {};
\end{scope}

\begin{scope}[xshift=4cm,yshift=-0.5cm,start chain=going below,node distance=7mm]
\foreach \i in {1,2,3}
  \node[ssnode,on chain] (s\i) [label=right: $\mathbf{a_{\i}}$] {};
\end{scope}

\node [myblue,fit=(f1) (f4),label=above:\textbf{Queries}, line width=.75mm] {};
\node [myred,fit=(s1) (s3),label=above:\textbf{Documents}, line width=.75mm] {};

\draw [line width=.75mm](f1) -- (s1);
\draw [line width=.75mm](s2) -- (f1);
\draw [line width=.75mm](f2) -- (s2);
\draw [line width=.75mm](s2) -- (f3);
\draw [line width=.75mm](s3) -- (f3);
\draw [line width=.75mm](f4) -- (s1);
\end{tikzpicture}
}
\caption{\label{fig1} Example bipartite graph induced by log data and document ranker.  Alternative queries for $q_1$ are $q_2, q_3$ if $a_2$ is logged and $q_4$ if $a_1$ is logged.}
\Description{A bipartite graph of queries and documents with edges connecting queries to documents returned by the query.}
\end{figure}
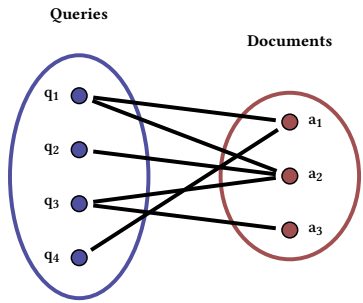

Specifically, the document ranker and the log data together induce a bipartite graph, with the queries on one side and the documents on the other, as illustrated in \cref{fig1}.  An edge exists between a query and a document if the document is returned by the ranker given the query.  We thus use the production ranker and the logged data in the retriever-training set to construct this bipartite graph and take as proposed alternative queries those that are neighbors of the logged document.  As an example, suppose that $q_1$ were recorded with the logged document being $a_1$.  Then the set of possible alternative queries would be $\{q_1, q_4\}$.  If instead $a_2$ were recorded as the logged document, the set of alternative queries would be $\{q_1, q_2, q_3\}$.  We make the retriever utility-aware by further filtering this set to include only queries with estimated utility at least 1.  In the example above, suppose that $\rank_{q_1}(a_1) = 5$, $\rank_{q_2}(a_1) = 10$ and $\rank_{q_3}(a_1) = 2$.  Then the set of alternative queries returned given a log entry of $(q_1, a_1)$ would be $\{q_1, q_3\}$, as $q_2$ has estimated utility less than one.  We then train the retriever using the prefixes as contexts and the proposed alternative queries as labels.

\begin{figure}[ht]
    \centering
    \scalebox{.7}{
    \begin{tikzpicture}[thick,
      every node/.style={draw,circle},
      fsnode/.style={fill=myblue},
      ssnode/.style={fill=myred},
      psnode/.style={fill=mygreen},
      every fit/.style={ellipse,draw,inner sep=-2pt,text width=2cm},
      ->,shorten >= 3pt,shorten <= 3pt
    ]
    
    \begin{scope}[yshift=-1cm,start chain=going below,node distance=7mm]
        \foreach \i in {1}
          \node[psnode,on chain] (p\i) [label=above: \textbf{Context}] {};
    \end{scope}

    \begin{scope}[xshift=4cm,yshift=-.5cm,start chain=going below,node distance=7mm]
    \foreach \i in {1,2}
      \node[fsnode,on chain] (f\i) [label=above: $\mathbf{q_\i}$] {};
    \end{scope}
    
    \begin{scope}[xshift=8cm,yshift=1cm,start chain=going below,node distance=7mm]
    \foreach \i in {1,2}
      \node[ssnode,on chain] (s\i) [label=right: $\mathbf{a_{\i}}$] {};
    \end{scope}
    
    \begin{scope}[xshift=8cm,yshift=-1.5cm,start chain=going below,node distance=7mm]
        \foreach \i in {3,4}
          \node[ssnode,on chain] (s\i) [label=right: $\mathbf{a_{\i}}$] {};
    \end{scope}

    \node [myblue,fit=(f1) (f2),label=above:\textbf{Proposed Queries}, line width=.75mm] {};
    \node [myred,fit=(s1) (s3),label=above:\textbf{Documents}, line width=.75mm] {};
    \node [myred,fit=(s3) (s4), line width=.75mm] {};

    \draw [line width=.75mm] (p1) -- (f1);
    \draw [line width=.75mm](p1) -- (f2);
    \draw [line width=.75mm](f1) -- (s1);
    \draw [line width=.75mm] (f1) -- (s2);
    \draw [line width=.75mm](f2) -- (s3);
    \draw [line width=.75mm](f2) -- (s4);
    \draw [line width=.75mm] (f1) -- (s3);
    
    \end{tikzpicture}
    }
    
    \caption{\label{fig2} Illustration of the construction of the data set used to train ranker.  The retriever returns the proposed queries given the context.  The production ranker is then used to find the documents given the retrieved queries.  This ranking is then used to calculate utilities.  Note that different queries can return the same document but in a different order.  For instance, $q_1$ ranks $a_3$ third while $q_2$ ranks $a_3$ first.}
    \Description{Illustration of a sample path from context to proposed queries to a ranking of documents given the proposed queries.}
    \end{figure}
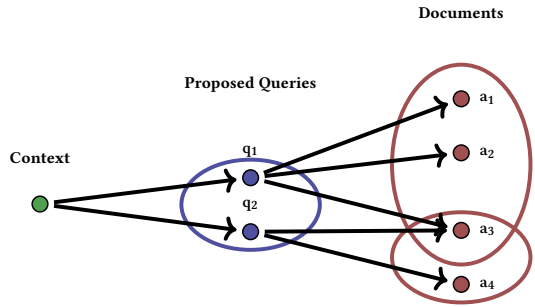

Now, given the trained context-to-query retrieval model, we turn to the second partition of the log data.  Using this retriever, we surface a list of proposed alternative queries for each log entry.  We use the production ranker to get the ranks of the logged document given the alternative queries; these ranks are then used to estimate the utility of the query.  Finally, we train a ranking model with these utilities as targets.  The process is illustrated in \cref{fig2}.

At inference time, for each logged entry, we use the query retriever followed by the ranker to return a ranked list of queries optimized for utility.  We can then use the production ranker to get the rank of the logged document in a query according to this policy and evaluate the average utility.


\section{Experimental Results}\label{sec:experiments}
We present detailed results on a benchmark eXtreme Multi-label learning dataset adapted to our setting (see \cref{app:conversion}), as well as demonstrating efficacy in a real world setting on a proprietary dataset from an online shopping store.
\subsection{Dataset Setup/Description}
\subsubsection{Amazon Titles Dataset}
We use the Amazon Titles dataset, collected by \cite{data1,data2} and made available as \textsf{LF-AmazonTitles-1.3M} on \cite{varma}.  This dataset has been collected by scraping the Amazon product catalogue.  In raw form, each datum consists of the title of an Amazon product together with some metadata, with labels consisting of other products suggested by Amazon on the web page of the first product.  The dataset consists of $1,305,265$ labels and $2,248,619$ training points, where again, each label is another Amazon product.  We filter and keep the top $50,000$ most frequent labels, thereby making the induced bipartite graph described above more dense.  We then train a PECOS model \cite{yu2020pecos} on the entire training set to serve as our document retrieval/product ranking system. After the product ranker is trained, we generate our logs according to a certain known click propensity model.  We restrict ourselves to the probability of a product observation being inversely proportional to the rank of the product given the query; a click is recorded if the observed product is relevant.  We created a log with $3,842,425$ samples using the above sampling procedure.  We then split the log using a $(.6, .3, .1)$ split on the instances, using the first set to train the candidate retriever (again a PECOS model), the second set to train the ranking model, and the last set to be used as a test set to validate our results.  To train the candidate retriever, we sampled prefixes by choosing a random truncation point after the first word in the title, and throwing away any prefixes that are too short.  We passed through the retriever training set $15$ times with this procedure, but only once for each of the ranker training and test sets.  For computational reasons, we then subsample the ranking train set by half.  The sizes of the data sets are summarized in \cref{tab:titles_stats}.
\begin{table}[t]
    \begin{center}
        \caption{\label{tab:titles_stats} Dataset statistics for \textsf{LF-AmazonTitles-1.3M} experiment.}
        \begin{tabular}{lccc}
            \toprule & Retriever Train & Ranker Train & Test \\ \midrule
            \# Samples & $25,981,965$ & $1,541,700$ & $386,648$ \\ 
            \# Distinct Contexts & $294,538$ & $147,411$ & $48,621$ \\ \bottomrule
        \end{tabular}
    \end{center}
\end{table}
We then use the retriever to generate a set of candidate queries associated to each prefix in the ranker training set and the test set.  In order to train a ranker, we need to featurize the prefixes and candidate queries.  To do this, we use a pre-trained model from the \textsf{transformers} library \cite{wolf-etal-2020-transformers}, specifically the \textsf{bert-base-uncased} model \cite{bert}, after normalizing the queries and prefixes to remove punctuation and make lower case.  We then use \textsf{XGBoost} \cite{Chen:2016:XST:2939672.2939785} with the PLTR objective to fit a ranker with the utilities as labels.  For all of our experiments, we used a depth 8 base classifier with 200 boosting rounds and all other parameters set to the default values.
\subsubsection{Data From Online Store}
There are some key differences in how we set up the experiments for data from an online store. First, we have no need to simulate log data and no need to train a document ranker as we are given logged data with all required information relating to the context, prefix, and document ranker. Second, we use extra contextual information beyond simply the prefix in order to suggest and rank queries.  Third, we featurize the contexts and proposed queries in a different way, relying on features developed internally by the online shopping store.  Other than this, the pipeline and evaluation are identical.  For instance, the candidate retriever we utilize is the PECOS model \cite{yu2020pecos} and is trained on the estimated utilities. The query re-ranker is a gradient boosted decision tree for which we use \textsf{XGBoost} \cite{Chen:2016:XST:2939672.2939785} with the PLTR objective to fit a ranker with the utilities as labels. The data are proprietary and we consider this section a demonstration of the verisimilitude of our model setting above, as well as positive evidence that the proposed system works in the real world. 

\subsection{Evaluation Metrics and Baselines}\label{sec:metricsBaselines}
We compare different QAC policies to the baseline logged query by assuming a click propensity model on the presented queries and reporting the \textbf{position-weighted (relative) utility} (denoted as \textsf{Utility@k}), defined as: 
\begin{align*}
\textsf{Utility@k} = \sum_{j=1}^k p_j \cdot \hat{u}(x,q_j|\bar{q},a),
\end{align*}
where, $\{q_j\}_{j=1}^k$ is a ranked list of queries returned by a ranking policy that is being evaluated, $\hat{u}(\cdot)$ is the utility estimator defined in \cref{eq:utility}, $x, a$ are respectively the context, a relevant document that the user clicks on by using the logged query $\bar{q}$.  We denote by $p_j$ the click propensity model associated to the \emph{queries}; for our metric, we consider $p_j \propto j^{-1}$.  Note that \textsf{Utility@1} is just the average utility of the top-ranked query according to the policy under consideration.  We also consider \textsf{Utility@5} and \textsf{Utility@10}, although we observe in our experiments that the relative quality of different QAC policies largely does not depend on which $k$ is used.  We remark that our notion of position-weighted utility comes directly from our model and corresponds to the factor increase of the probability of finding a relevant document under the click propensity model.  Alternatives such as Mean Recipricol Rank (MRR) and Discounted Cumulative Gain (DCG) are not well-suited to measure the utility; thus, while the utility-aware ranker may be less competitive on these metrics, they do not naturally fit into the current framework.  As such, we restrict our focus to the central theme of the paper: downstream utility.

We compare the proposed approach against a variety of baselines:
\begin{itemize}
    \item \textsf{Unbiased}: retrieval following by re-ranking system with utilities estimated using \cref{eq:utility}.
    \item \textsf{PECOS}: retrieval system's performance when trained with utilities estimated using \cref{eq:utility}. This serves to highlight the impact of the re-ranking step.
    \item \textsf{Oracle}: evaluates a ranking policy that returns the best query in terms of utility, which requires knowing utility for test data and is thus not implementable. This serves to highlight the gap one can hope to bridge by developing more powerful classes of re-ranking algorithms.  Note that only \textsf{Utility@1} is meaningful for this policy.
    \item \textsf{Logged}: The (relative) utility of the logged query.  Note that this is the baseline to which the other policies are compared.  By \cref{eq:utility}, the logged queries always have relative utility $1.0$.  Note that only \textsf{Utility@1} is meaningful for this policy.  
    \item \textsf{Random}: Performance of a policy that returns queries retrieved by the PECOS retriever in a uniformly random order.
\end{itemize}
\subsection{Empirical Results - Amazon Titles Dataset}
In this section we describe our results on the Amazon Titles data.  In \cref{sec:core_results} we compare our approach to the baselines described in \cref{sec:metricsBaselines}; in \cref{sec:increasingdata} we explore the how increasing the amount of data affects the quality of our ranker; in \cref{sssec:debias_effect} we compare our proposed utility estimator from \cref{eq:utility} to other natural alternatives; and in \cref{sssec:mis-specification} we demonstrate that our approach is robust to misspecification of the click propensity model.

\subsubsection{Core Results}\label{sec:core_results}
\begin{figure}[!htb]
    \centering
    \includegraphics[width=\linewidth]{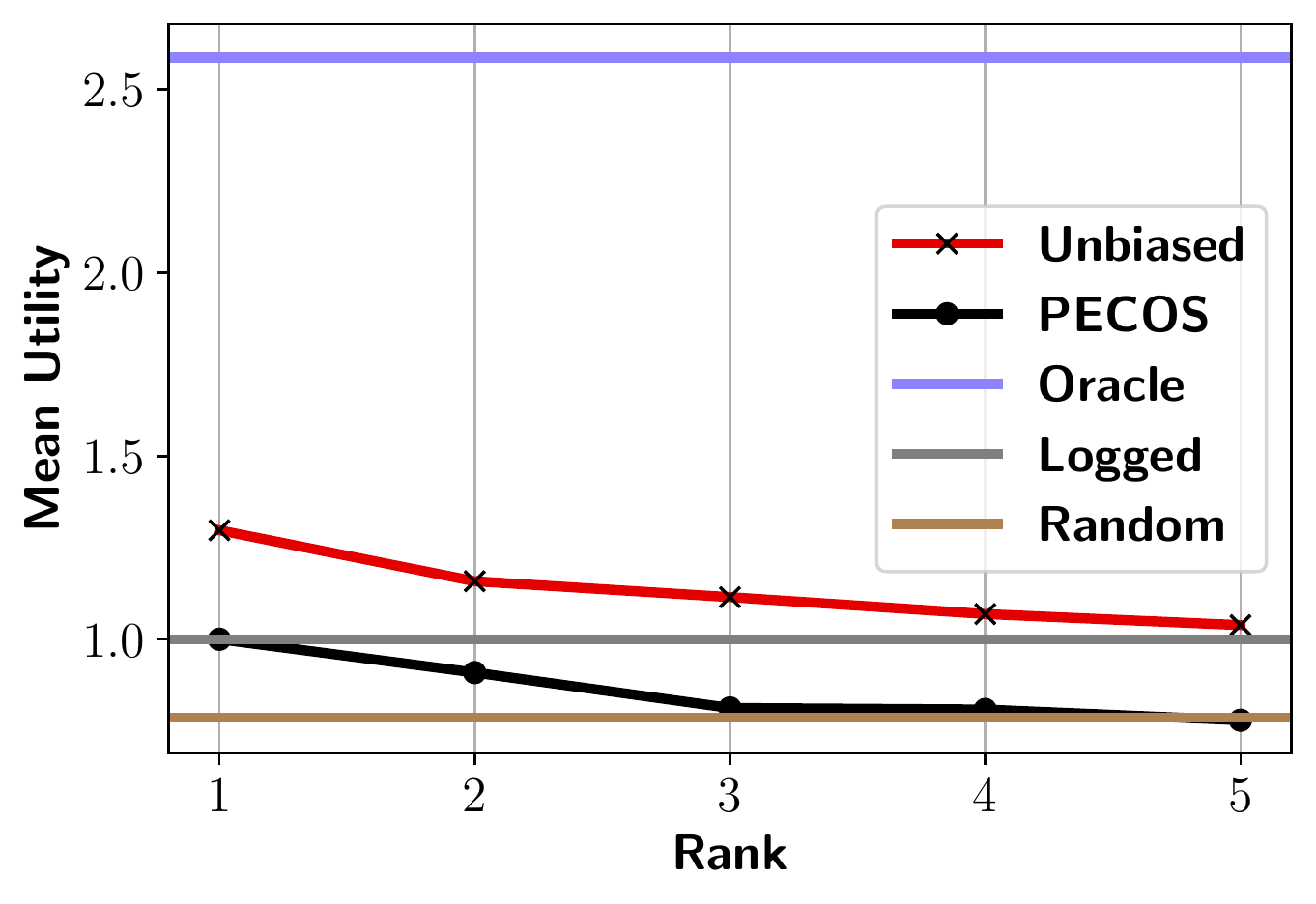}
    \caption{Average utilities at each position according to different policies.  Note the utilitiy-aware Ranker performs best.}\label{fig:utility_and_rank}
    \Description{Illustration comparing mean utilities at ranks 1 through 5 of different policies.  Our proposed approach does better than the logged baseline, which in turn beats the PECOS model, which in turn beats the random policy.}
\end{figure}

\cref{tab:ablation_stats} presents the core results comparing the proposed approach against various baselines presented in \cref{sec:metricsBaselines}.  We also plot the average utility at each position $k$ for $1 \leq k \leq 5$, according to each policy in \cref{fig:utility_and_rank}.  Taken together, these results suggest that the proposed QAC framework significantly outperforms the benchmark \textsf{Logged} policy in addition to policies that only rely on the retriever (\textsf{PECOS} and \textsf{Random}).  Note that the proposed re-ranker has the potential to be improved upon since the (indadmissable) oracle strategy results in still better queries.
\begin{table}[t]
    \begin{center}
        \caption{\label{tab:ablation_stats} Comparing Position-Weighted Utilities of the proposed framework against core baseline methods outlined in \cref{sec:metricsBaselines}. See \cref{sec:core_results} for more details.}
        \begin{tabular}{lccc}
            \toprule Ranking Policy & Utility@1 & Utility@5 & Utility@10 \\ \midrule
            \textsf{Logged} & $1.000$ & - & - \\             
            \textsf{PECOS} & $1.000$ & $.9128$ & $.8692$ \\ 
            \textsf{Oracle} & $2.587$ & - & - \\
            \textsf{Random} & $.7867$ & $.7867$& $.7867$\\ \midrule
            \textsf{Unbiased (\textbf{proposed})} & $\mathbf{1.297}$ & $\mathbf{1.232}$ & $\mathbf{1.185}$ \\ 
            \bottomrule
        \end{tabular}
    \end{center}
\end{table}

\subsubsection{Does more Data lead to a Better Ranker?}\label{sec:increasingdata}
\begin{figure}[!htb]
    \centering
     \includegraphics[width=\linewidth]{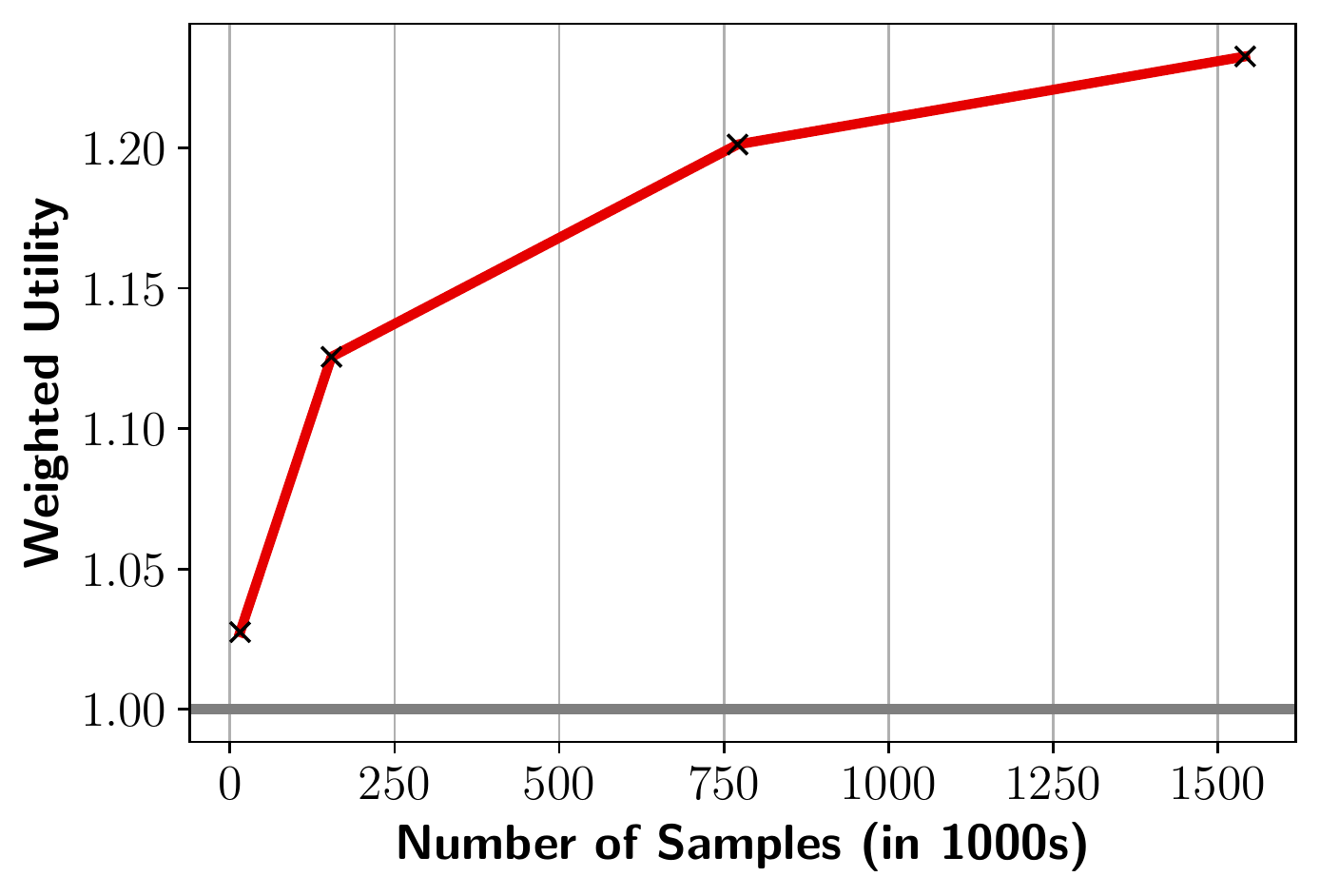}
     \caption{Position Weighted Utilities truncated at 5 according to the utilitiy-aware Ranker trained on samples of different sizes.}\label{fig:size_ablation5}
     \Description{Illustration of mean weighted utility to rank 5 as the number of training samples grows from under 100,000 to over 1.5 million.  The utility follows the expected concave, increasing function the theory predicts.}
\end{figure}

We examine the effect that increasing the size of the data set has on the quality of the utility-aware ranker, summarized in \cref{fig:size_ablation5}.  

We subsample the log data at sizes $1\%$, $10\%$, and $50\%$ and train the utility-aware ranker on each of these data sets, demonstrating a clear upward trend.  While we evaluate under the Position-Weighted Utility truncated at 5 here, utility at 1 and 10 exhibit similar behavior, as can be seen in \cref{app:figs}.  As predicted from \cref{thm:generalization}, the performance of the utility-aware ranker increases with the sample size.  Note that while the utility still does not approach the quality of \textsf{Oracle} even with more than 1.6 million samples, it is not clear that the \textsf{Oracle} policy is even included in the set of policies achievable by our base classifier.

\begin{figure}[!htb]
    \centering
     \includegraphics[width=\linewidth]{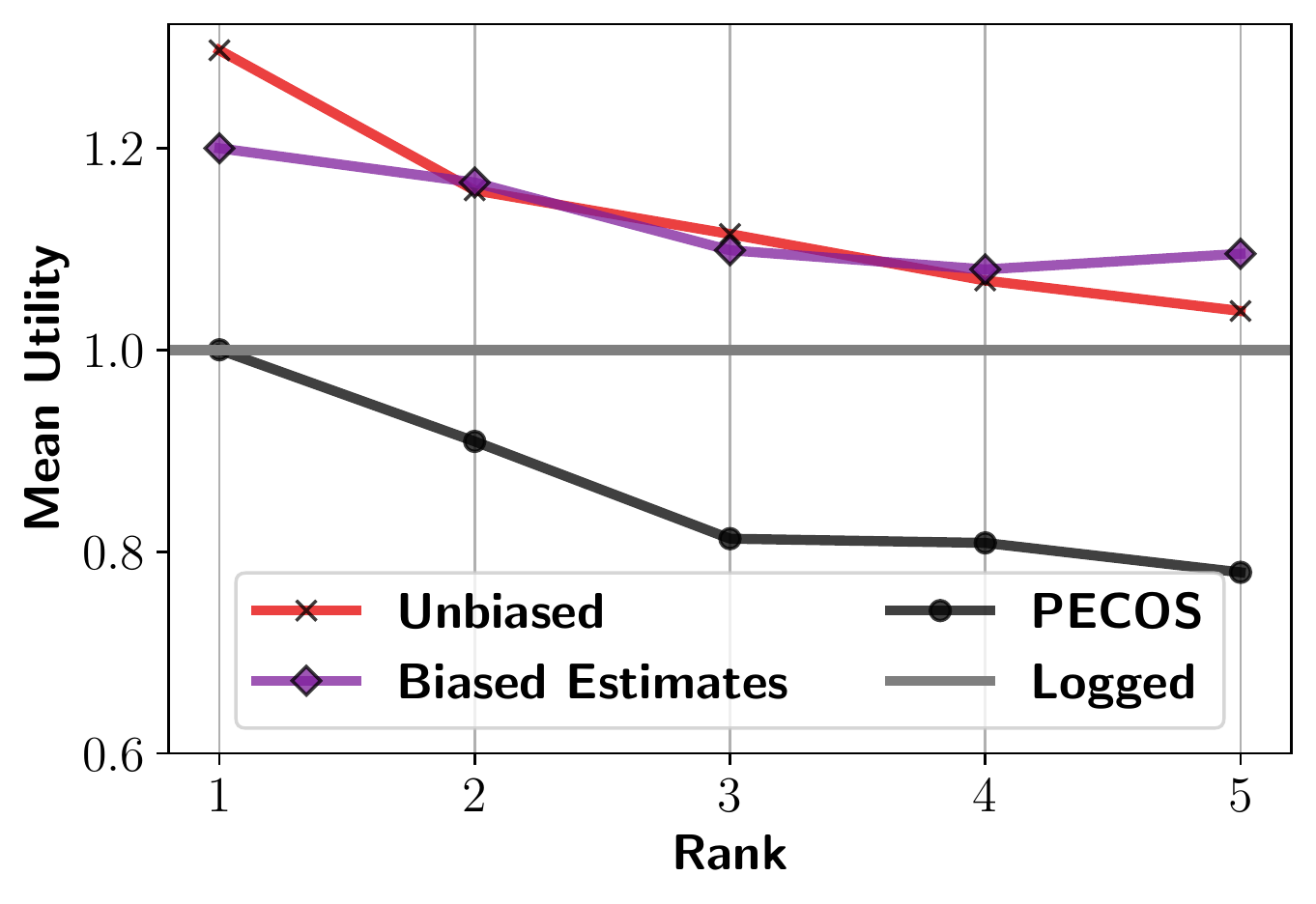}
     \caption{Utilities at each position according to different policies.  \textsf{Biased Estimates} is trained on utility estimates not adjusted using the inverse propensity score.}\label{fig:biased}
     \Description{Illustration of mean utility at positions up to 5, demonstrating that the proposed policy performs better than one that forgoes the debiasing based on click propensities.}
\end{figure}

\begin{figure}[!htb]
    \centering
     \includegraphics[width=\linewidth]{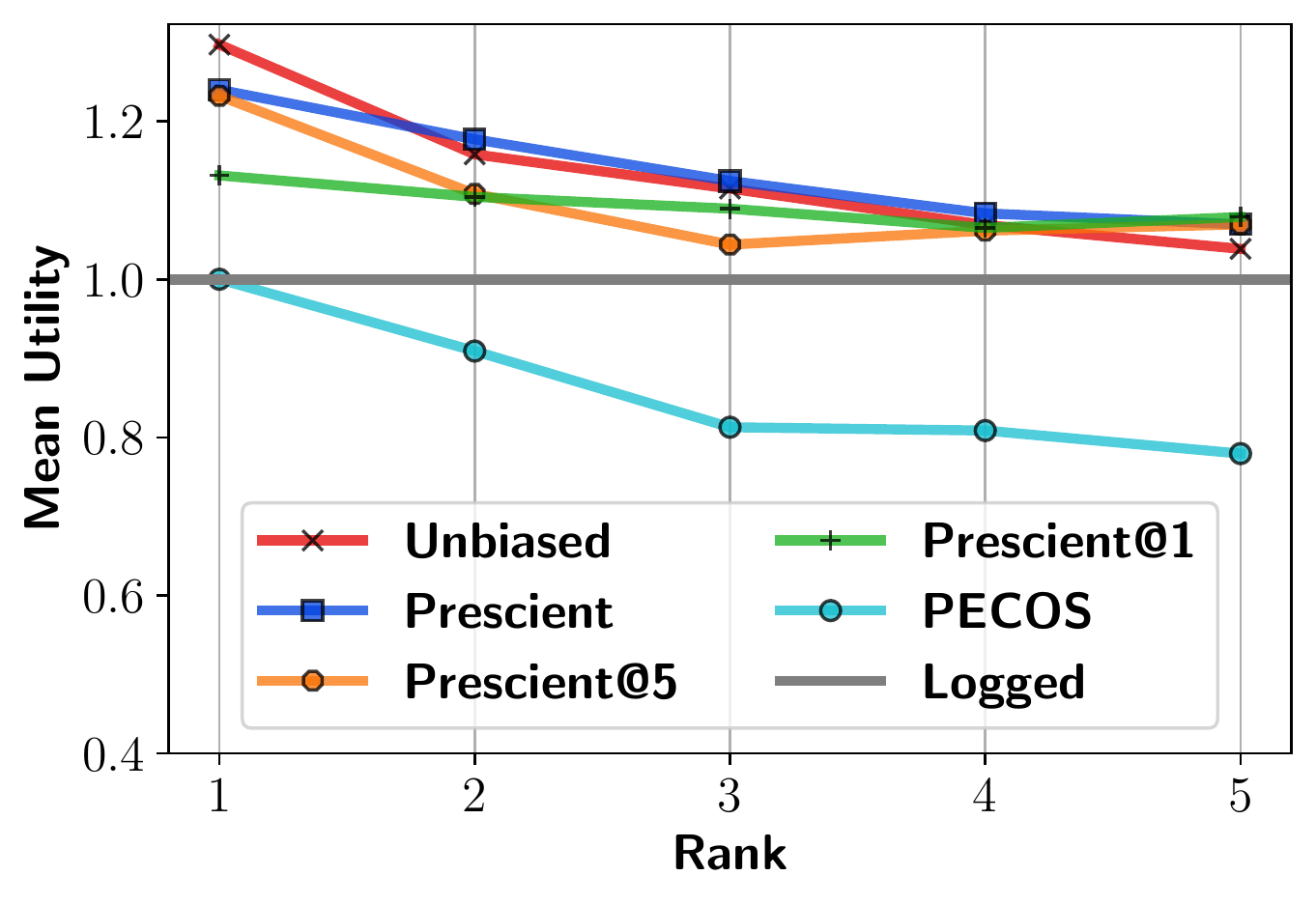}
     \caption{Utilities at each position according to different policies.  Note that prescience hurts performance relative to the unbiased utility estimates.}\label{fig:prescient}
     \Description{Illustration of mean utility at positions up to 5, demonstrating that the prescient ablation described in text performs worse than the proposed unbiased policy.}
\end{figure}

\subsubsection{What is the impact of de-biased utility estimator \cref{eq:utility} over other biased alternatives?}\label{sssec:debias_effect}
We consider what happens when we train the ranker to use utility estimates other than our unbiased estimator.  In this series of ablation experiments, at each step we take away more information available to the ranker and demonstrate the necessity of some of our assumptions.  The performance of these ablations is summarized in \cref{fig:biased} with metrics in \cref{tab:ablation_stats_1}.  For each of these experiments, with notation from \cref{as4}, we observe a log entry $(x, \overline{q}, a, \rank_{\overline{q}}(a))$ where $x$ is the prefix, $\overline{q}$ is the logged query, and $a$ is the logged document.  We change only how we estimate the quality of a new query $q$ with respect to the prefix $x$.  We begin by considering a ranker trained by neglecting the biased nature of clicks in the logged data, described as:
\begin{itemize}
	\item \textsf{Biased Estimates}, trained on utility estimates that are not de-biased by multiplying by an inverse propensity score.  In particular, we train our ranker on targets $p_{\rank_{q}(a)}$
    , which, due to the nature of log collection, amount to biased estimates of the objective.  
\end{itemize}
As expected (refer to \cref{tab:ablation_stats_1}), removing the click propensity correction adds bias to the estimates and hurts the quality of the final ranker. 
\begin{table}[t]
    \begin{center}
        \caption{\label{tab:ablation_stats_1} Comparing Position-Weighted Utilities of proposed framework (with un-biased utility estimates) against biased utility estimators described in \cref{sssec:debias_effect}.}
        \begin{tabular}{lccc}
            \toprule Ranking Policy & Utility@1 & Utility@5 & Utility@10 \\ \midrule
            \textsf{Biased Estimates} & $1.200$ & $1.176$ & $1.146$ \\
            \textsf{Prescient} & $1.240$ & $1.222$ & $1.178$\\
            \textsf{Prescient@5} & $1.233$ & $1.179$ & $1.131$\\
            \textsf{Prescient@1} & $1.131$ & $1.104$ & $1.076$ \\\midrule
            \textsf{Unbiased (\textbf{proposed})} & $\mathbf{1.297}$ & $\mathbf{1.232}$ & $\mathbf{1.185}$ \\ 
            \bottomrule
        \end{tabular}
    \end{center}
\end{table}
Second, we try to model the user experience, where a user who knows for which document she is searching can reason about likely queries; because users tend to not explore deep into the ranking, we suppose that the user is only able to reason correctly about queries leading to a desired document ranked sufficiently highly.  We refer to this policy as \emph{prescience}, the ability of the ranker to see only whether or not a relevant document is ranked by the query in a high position, but without knowledge of the precise position of the relevant document.  Formally, for fixed $k \in \mathbb{N} \cup \{\infty\}$, we consider:  
\begin{itemize}
    \item \textsf{Prescient@k}, trained on utility estimates $\mathbf{1}[\rank_q(a) \leq k]$, assigning positive utility to queries ranking a relevant document sufficiently high.
\end{itemize}
We train these policies for $k \in \{1,5, \infty\}$, denoted by \textsf{Prescient@1}, \textsf{Prescient@5}, and \textsf{Prescient}.  The average utilities at each rank are summarized in \cref{fig:prescient} with metrics in \cref{tab:ablation_stats_1}.  Unsurprisingly, the ranker trained on the unbiased estimates does better than any of the prescient policies and the quality of the \textsf{Prescient@k} models decreases along with $k$.  Note that this second fact is not obvious as there are two competing factors contributing information to the \textsf{Prescient@k} data: first, with a higher $k$, the training data provide strictly more information about which queries contain documents relevant to which contexts (with $k = \infty$ simply binarizing our estimated utility); second, with a lower $k$, the training data provide more \emph{positional} information about which queries rank relevant documents more highly.  Thus, there are two competing effects, although the first effect clearly dominates in our experiments.  

Somewhat surprisingly, \textsf{Prescient} is better than \textsf{Biased Estimates}.  We suspect that this is an artifact of the specific PLTR reduction that \textsf{xgboost} uses, where, without instance weighing, the magnitude of the utility is irrelevant.  Unsurprisingly, as $k$ decreases, the quality of the \textsf{Prescient@k} policy tends in the same direction, corresponding to the fact that less information is available to the ranker.
\begin{figure}[t]
    \centering
    \includegraphics[width=.45\textwidth]{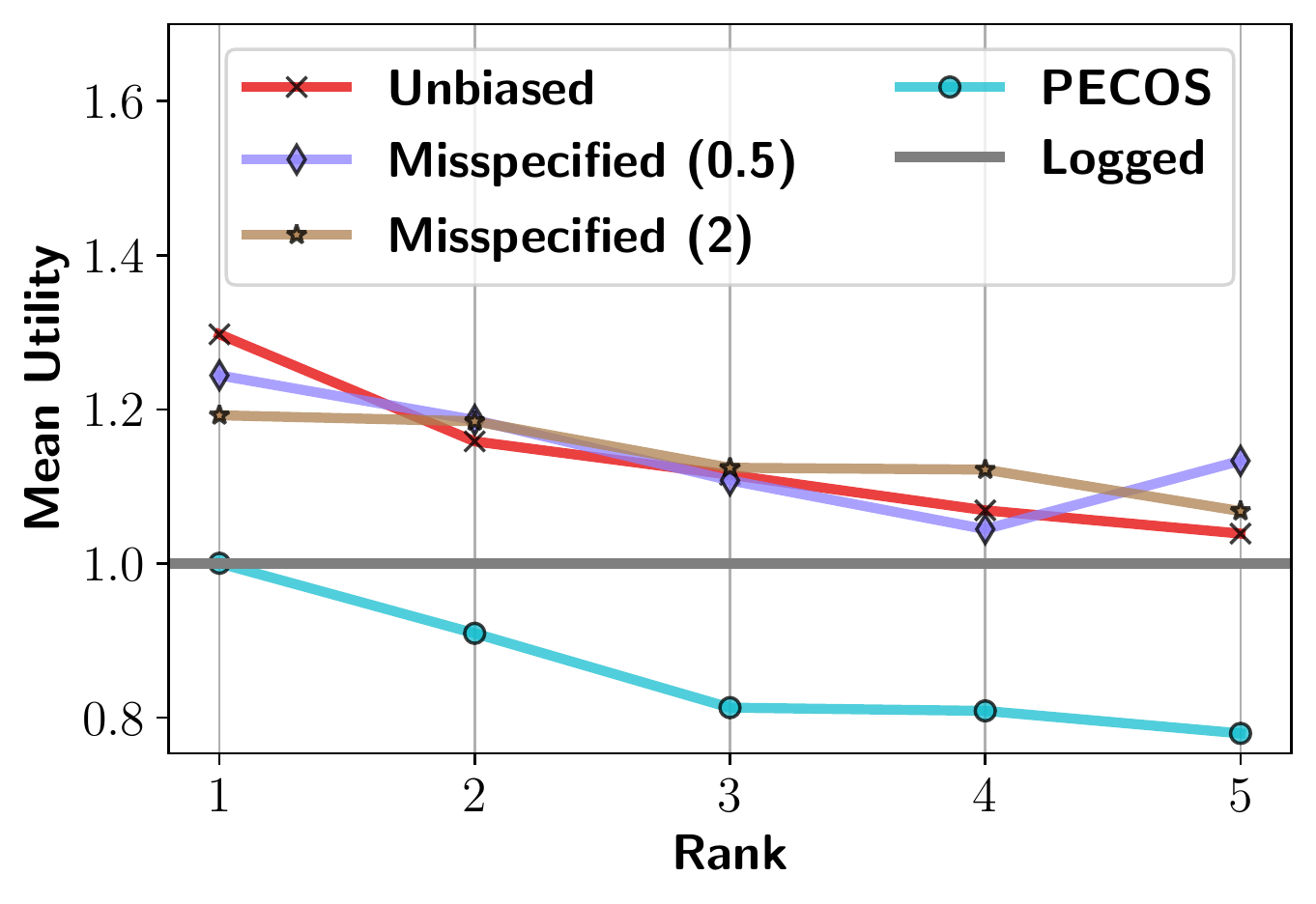}
    \caption{\label{fig:propensity} Average utilities at each position according to different policies.  \textsf{Misspecified (0.5)} is the ranker trained on utility estimates assuming $p_i \propto i^{- \frac 12}$, \textsf{Misspecified (2)} is the ranker trained on utility estimates assuming $p_i \propto i^{- 2}$.}
    \Description{Illustration of mean utility at positions up to 5 demonstrating the effect of misspecifyingthe click propensity model.  In particular, while the correct model does best, there is limited negative effect if the tails of the click model are overestimated, but a much more pronounced decline of performance when the tails are underestimated.}
\end{figure}
\subsubsection{How robust is the proposed approach to mis-specification of click propensities?}\label{sssec:mis-specification}
\begin{table}[t]
    \begin{center}
        \caption{\label{tab:ablation_stats_2} Position-Weighted Utilities of the proposed strategy when trained with mis-specified click propensity models. Despite mis-specification, it is worthwhile noting that each of these estimators still outperforms the logging policy that has a relative utility of $1.0$.}
        \begin{tabular}{lccc}
            \toprule Ranking Policy & Utility@1 & Utility@5 & Utility@10 \\ \midrule
            \textsf{Propensity0.5} & $1.244$ & $1.221$ & $1.174$ \\
            \textsf{Propensity2} & $1.192$ & $1.158$ & $1.128$ \\\midrule
            \textsf{Unbiased (\textbf{proposed})} & $\mathbf{1.297}$ & $\mathbf{1.232}$ & $\mathbf{1.185}$ \\ 
            \bottomrule
        \end{tabular}
    \end{center}
\end{table}
We also consider what happens when we mis-specify the click propensity model.  Having simulated the logs ourselves, we know that the probability of a click is inversely proportional to the rank.  In this experiment, we keep the logs the same, but estimate the utility with a different click propensity model:
\begin{itemize}
    \item For a fixed $\alpha > 0$, denote by \textsf{Misspecified}$(\alpha)$ a utility-aware ranker trained using the click propensity model $p_k \propto k^{- \alpha}$.
\end{itemize}
We try $\alpha \in \{.5, 2\}$ and refer to these models as \textsf{Misspecified (0.5)} and \textsf{Misspecified (2)}.  The average utilities against ranks are summarized in \cref{fig:propensity} with metrics included in \cref{tab:ablation_stats_2}.  Once again, we see that our utility-estimate-trained ranker outperforms the misspecified ranker, although the misspecification does not destroy performance.  Interestingly, mis-specification is more tolerable if we make the tails fatter than if we shrink the tails.  This is unsurprising given \cref{prop:variance}, which says that the variance of the estimator depends on the maximum of the restricted likelihood ratio; the fatter the tail, the lower this bound.  Thus, while mis-specifying the click propensity model introduces bias, if we over-fatten the tails we can also significantly reduce variance, as observed in \cite{joachims2017unbiased}.
\subsection{Empirical Results - Data from an Online Shopping Store}
Considering data from a real world online shopping store, we present an empirical validation of the utility-aware QAC system by comparing our utility-aware estimation procedure against baselines described in \cref{sec:metricsBaselines}; in particular, we consider the utility-aware ranker (\textsf{Unbiased}), the retriever (\textsf{PECOS}), the \textsf{Oracle}, and the \textsf{Random} policies.  Our results are presented in table \cref{tab:amazon_stats}.
\begin{table}[t]
        \caption{\label{tab:amazon_stats} Performance of different ranking policies on data from an online shopping store.}
        \begin{tabular}{cccccc}
            \toprule
            \begin{tabular}[x]{@{}c@{}}Ranking\\Policy\end{tabular} & Logged & \textsf{Unbiased} & \textsf{PECOS} & \textsf{Oracle} & \textsf{Random}\\
            \midrule
            Utility@1 & 1.0 & $\mathbf{1.979}$ & $1.535$ & $3.890$ & $1.290$\\
            \bottomrule
        \end{tabular}
\end{table}
Once again we see a clear advantage of the proposed approach over the \textsf{PECOS} model.  The primary difference between this setting and that of the simulated data above is that the utilities are much higher.  In fact, we believe that the retriever is much better at surfacing high-utility queries both due to the added contextual information and the volume of data on which each model was trained.

\section{Conclusion}
In this work, we consider a utility maximizing perspective to optimizing QAC systems and cast it as one of optimizing a ranking of rankings task. We then present an approach to learn such a ranking policy given logs of biased clickthrough data by proposing an unbiased utility estimator, bounding its variance and generalization error of the policy under standard learning theoretic conditions. We present experimental results on simulated data and real-world clickthrough QAC logs. We present ablation studies that include how the method performs with a mis-specified click propensity model. Questions for future work include how we can go beyond additive utilities, such as considering sub-modular utility functions.

\begin{acks}
    AB acknowledges support from the National Science Foundation Graduate Research Fellowship under Grant No. 1122374.
\end{acks}

\bibliographystyle{ACM-Reference-Format}
\bibliography{references}
\appendix
\section{Proofs}\label{app:proofs}
\begin{proof}[Proof of \cref{prop:unbiased}]
    Unravelling definitions, we have:
    \begin{align}
        \widehat{u}(x, q| \qbar, a)  = \sum_{\substack{a \in q \\ r(x, a) = 1}} \frac{p_{\rank_q(a)}}{p_{\rank_{\qbar}(a)}} o(\qbar, a)
    \end{align}
    Taking expectations with respect to $o(\qbar, a)$ yields utility $u(x,q)$.
\end{proof}

\begin{proof}[Proof of \cref{prop:novariancecontrol}]
    Note, first, that
    \begin{align}
        &\variance(\widehat{u}(x, q| \qbar, a)) =  \\
        &\ee\left[\sum_{\substack{a \in \qbar \\ r(x, a) = 1}} \frac{\pqa p_{\rank_q(a')}}{p_{\rank_{\qbar}(a')}} o(\qbar, a) o(\qbar, a') \right] - u(x, q)^2 \label{eq:var1} \\
        &= \sum_{\substack{a \in \qbar \\ r(x, a) = 1}} \frac{\pqa^2}{\pqba} - u(x, q)^2 \label{eq:var2}
    \end{align}
    where \cref{eq:var1} follows from expanding the definition of $\widehat{u}$ and the fact that it is an unbiased estimator and \cref{eq:var2} follows from the fact that $o(\qbar, a) o(\qbar, a') = o(\qbar, a) \delta_{aa'}$ with $\delta_{aa'}$ the Kronecker $\delta$.  Suppose there is a unique $a \in q \cap q'$; following from eq:~\ref{eq:var2}:
    \begin{align}
        \variance(\widehat{u}(x, q | \qbar ,a)) &= \frac{\pqa^2}{\pqba^2} - \pqa^2 \geq \frac{\pqa^2}{\pqba^2} - 1
    \end{align}
    Letting $\frac{\pqa^2}{\pqba^2} = C + 1$ concludes the proof.

\end{proof}

\begin{proof}[Proof of \cref{prop:variance}]
    The first statement is clear from the definition of utility.  The second statement follows immediately from \cref{eq:var2}.
\end{proof}

\begin{proof}[Proof of \cref{lem:losses}]
    This follows from the tower property of conditional expectation, linearity, and \cref{prop:unbiased}.
\end{proof}

\begin{proof}[Proof of \cref{thm:generalization}]
    By \cref{lem:losses}, it suffices to consider $\widetilde{L}$ instead of $L$.  From learning theory (see \cite{wainwright2019high,devroye2013probabilistic}) we know that
    \begin{equation}
        \ee\left[\widetilde{L}(S_n) - \widetilde{L}(S^*)\right] \leq 2 \ee\left[\sup_{S \in \F} \widehat{L}_n(S) - \widetilde{L}(S)\right]
    \end{equation}
    We now use the classical symmetrization technique to control the right hand side by $\R_n(\F)$.  
    Let
    \begin{equation}
        \Delta_{q,q'}(x) = \widehat{u}(x_i, q| \qbar_i, a_i) - \widehat{u}(x_i, q'| \qbar_i, a_i)
    \end{equation}
    and $I_{q,q'}(x)$ be the event that $q, q' \in Q(x)$.  Furthermore, let $\ee_{|X}$ denote expectation conditional on the set of $x_1, \dots, x_n$.  
    Then a standard symmetrization approach yields:
    \begin{align}
        &\ee_{|X}\left[\sup_{S \in \F} \widehat{L}_n(S) - \widetilde{L}(S)\right] \\
        &\leq 2 \ee_{|X}\left[\sup_{S \in \F} \frac 1n \sum_{i = 1}^n \epsilon_i\frac 1{\binom{K}{2}}\sum_{q, q'} \Delta_{q,q'}(x) I_{q,q'}(x)S(x_i, q, q') \right]\\
        &\stackrel{(a)}{\leq}  \frac 1{\binom{K}{2}}\sum_{q, q'} \ee_{|X}\left[\sup_{S \in \F} \frac 1n \sum_{i = 1}^n \epsilon_i \Delta_{q,q'}(x) (2I_{q,q'}(x))S(x_i, q, q') \right]\\
        &\stackrel{(b)}{\leq} \frac 1{\binom{K}{2}}\sum_{q, q'} \ee_{|X}\begin{bmatrix}\sup_{S \in \F} \frac 1n \sum_{i = 1}^n \epsilon_i \Delta_{q,q'}(x) S(x_i, q, q') (2 I_{q,q'}(x) - 1) \\+  \Delta_{q,q'}(x) S(x_i, q, q')\end{bmatrix} \\
        &\stackrel{(c)}{\leq} \frac{2}{\binom{K}{2}} \sum_{q, q'} \ee_{|X}\left[\sup_{S \in \F} \frac 1n \sum_{i = 1}^n \epsilon_i \Delta_{q,q'}(x) S(x_i, q, q')\right]\\
        &\stackrel{(d)}{\leq} \frac{2 B}{\binom{K}{2}} \sum_{q, q'} \ee_{|X}\left[\sup_{S \in \F} \frac 1n \sum_{i = 1}^n \epsilon_i S(x_i, q, q')\right]
    \end{align}
    Where (a) follows since supremum of a sum is controlled by sum of suprema, (b) follows from positive homogeneity of expectation and supremum, (c) owing to independence of $\epsilon_i$, and because these are equal in distribution to the collection of $\epsilon_i (2 I_{q,q'}(x) - 1)$, and (d) follows by contraction and because \cref{prop:variance} implies that, as $0 \leq \widehat{u} \leq B$, $\abs{\Delta}$ satisfies the same bound.  Finally, in the outside sum, the $q, q' \in \bigcup_i Q(x_i)$.  We ignore queries whose utilities are zero because $\widehat{u}$ is also zero.  Thus number of terms in the sum is bounded by second binomial coefficient of number of queries relevant to at least one $x_i$.  Taking expectations 
     concludes the proof.
\end{proof}

\section{Simulating Log Data from XMC Data}\label{app:conversion}
We describe our conversion of XMC dataset into the utility-aware ranking problem. To demonstrate our results, we require a dataset with (1) raw text queries, (2) a document retrieval system, and (3) sufficient density in the induced bipartite graph between queries and documents.  We need clickthrough logs including queries, relevant documents, and their ranks given a logged query.  We consider the XMC dataset as consisting of contexts in the form of raw text and a set of labels associated to that context.  We view the raw text as queries and labels as documents.  It is clear that (1) holds; furthermore, (3) is a feature of the specific data set.  To obtain (2), we train a PECOS model \cite{yu2020pecos} where the contexts are queries and labels are the documents.  This PECOS model is treated as the master document ranker that returns ranked list of products given a query.

We generate log data as follows.  We sample a query $q$ from the dataset.  We use the PECOS model to produce ranked list of items $\rank_q$.  We use a click propensity model where $\pp(o(a) = 1) \propto \rank_q(a)^{- 1}$ for each $a$ returned by the ranker.  If $a$ is relevant for the query $q$, then we record $q$, $a$, and $\rank_q(a)$.  Otherwise, we throw out the sample and repeat.  To get the rank of the relevant product $a$ given a different query $q'$, we can simply find $a$ in $\rank_{q'}$.  For each entry, we randomly truncate the logged query anywhere after the end of the first word and record the resulting prefix. The log data thus consists of a prefix, a query, a product, and a position.

The above approach is justified in that it is actually fairly similar to the way in which many production product catalogs are constructed.  While the production ranker may use more features and consider a more complicated, business-oriented objective, the basic principle remains the same.  Similarly, our generation of log data is reasonably close to that which occurs in practice, with the caveat that the click propensity model is likely more involved in practice.  The prefixes to queries are often recorded in practice, but, in order to increase the size of the data set, random truncation is also a reasonable strategy and we use it in our implementation on the real data from an online shopping store as well.

\section{Extra Figures}\label{app:figs}
In this appendix, we compile figures relevant to our experiments for which we did not have space in the main text, in particular the effect of increasing the training set size on the other two weighted utility metrics.
\begin{figure}[!ht]
    \centering
    \includegraphics[width=.4\textwidth]{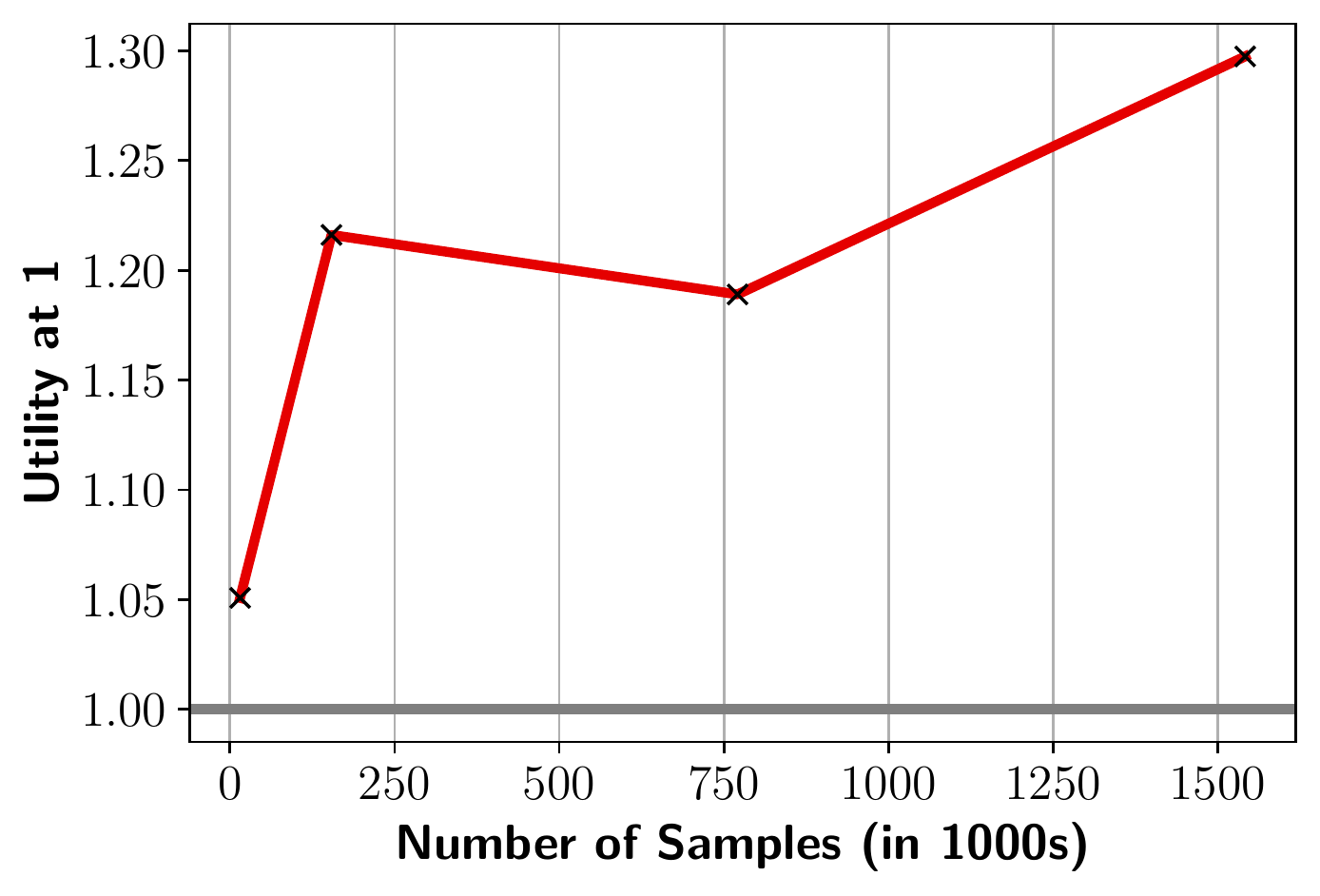}
    \caption{\label{fig:size_ablation1} Average utility of first ranked query according to the Utility-Aware Ranker trained on samples of different sizes.}
    \Description{Illustration of mean weighted utility at top rank as the number of training samples grows from under 100,000 to over 1.5 million.  The utility follows the expected concave, increasing function the theory predicts.}
\end{figure}

\begin{figure}[!ht]
    \centering
    \includegraphics[width=.4\textwidth]{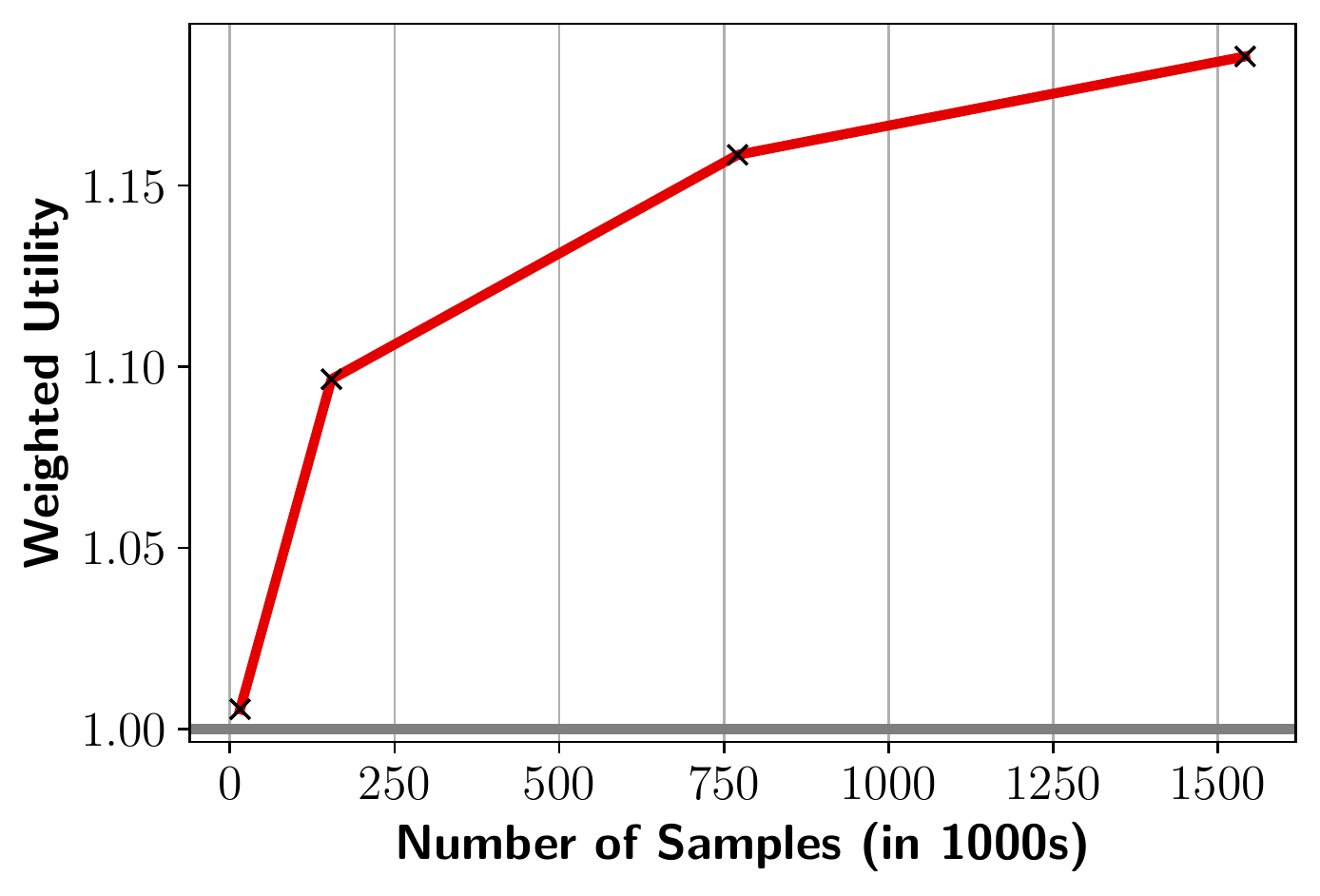}
    \caption{\label{fig:size_ablation10} Position Weighted Utilities truncated at 10 of the Utility-Aware Ranker trained on samples of different sizes.}
    \Description{Illustration of mean weighted utility to rank 10 as the number of training samples grows from under 100,000 to over 1.5 million.  The utility follows the expected concave, increasing function the theory predicts.}
\end{figure}

\end{document}